# Large magnetoresistance and spin-dependent output voltage in a lateral MnGa/GaAs/MnGa spin-valve device

Koki Chonan[1], Nguyen Huynh Duy Khang[1], Masaaki Tanaka[2,3], and Pham Nam Hai[1,3]

[1] *Department of Electrical and Electronic Engineering, Tokyo Institute of Technology*
*2-12-1 Ookayama, Meguro, Tokyo 152-8850, Japan*

[2] *Department of Electrical Engineering and Information Systems, The University of Tokyo,*
*7-3-1 Hongo, Bunkyo, Tokyo 113-8656, Japan*

[3] *Center for Spintronics Research Network (CSRN), Graduate School of Engineering,*
*The University of Tokyo, 7-3-1 Hongo, Bunkyo, Tokyo 113-8656, Japan*

**Abstract**

**We investigated the spin-dependent transport properties of a lateral spin-valve device with a 600 nm-long GaAs channel and ferromagnetic MnGa electrodes with perpendicular magnetization. Its current-voltage characteristics show nonlinear behavior below 50 K, indicating that tunnel transport through the MnGa/GaAs Schottky barrier is dominant at low temperatures. We observed clear magnetoresistance (MR) ratio up to 12% at 4 K when applying a magnetic field perpendicular to the film plane. Furthermore, a large spin-dependent output voltage of 33 mV is obtained. These values are the highest in lateral ferromagnetic metal / semiconductor / ferromagnetic metal spin-valve devices.**

# 1. Introduction

Spin transistors whose source and drain are made of ferromagnetic materials, such as the spin metal-oxide-semiconductor field-effect-transistor (spin MOSFET),[1,2] are promising three-terminal devices with embedded non-volatile memory capability. In spin transistors, the drain current depends not only on the gate voltage but also on the relative magnetization configuration of the source and drain. With this spin-dependent output characteristics, spin transistors can be applied to non-volatile memory cells[3] and reconfigurable logic circuits.[4] The operation of spin transistors requires highly efficient spin injection into a semiconductor (SC) and detection by ferromagnetic metal (FM) electrodes. However, due to the conductivity mismatch between the FM electrodes and SC channel in the diffusive transport regime, the imbalance of injected majority and minority spins in the SC channel becomes extremely small. In fact, very small magnetoresistance (MR) ratio (< 1%) and spin-dependent output voltage below 1 mV have been reported in previous study of micron-scale spin MOSFETs.[5-9] Recently, large MR ratios were reported in lateral spin-valve devices using the ferromagnetic semiconductor GaMnAs as source and drain electrodes.[10,11] Although using ferromagnetic semiconductor GaMnAs helps to suppress the conductivity mismatch, practical applications seem difficult due to its low Curie temperature (≤ 200 K).[12] Another pathway to overcome the conductivity mismatch problem is to use ballistic transport in nanoscale SC channels.[13,14] In recent studies of Si-based nanoscale spin-valve devices with Fe/MgO/Ge spin injectors and detectors and 20 nm Si channels, we obtained large spin-valve ratios up to 3.6% and large spin-dependent output voltages up to 25 mV.[15,16,17] Here, the spin-valve ratio is given by ($R_{AP}$-$R_P$)/$R_P$, where $R_{AP}$ and $R_P$ are the resistance of the spin-valve device in antiparallel and parallel magnetization configuration, respectively. Further performance improvement of Si-based nanoscale spin-valves is difficult since it is technically challenging to shorten the Si channel and grow high quality thin MgO layer on Si substrates. To improve the MR ratio and spin-dependent output voltage for practical applications, it is essential to choose suitable ferromagnetic metal and semiconductor materials with small lattice mismatch for epitaxial growth. Indeed, a relatively large MR ratio of 3.6% was reported in a lateral epitaxial MnAs/GaAs/MnAs spin-valve device with a 0.5 μm channel.[18] Even larger MR ratios up to 8.2% were observed in vertical epitaxial MnAs/10-30 nm GaAs/MnAs spin-valves.[19] However, the MnAs electrodes have in-plane magnetization, which is not suitable for miniaturization. In this aspect, the MnGa alloy with $L1_0$ ordering is a potential candidate for the source and drain electrodes. Good homogeneity and abrupt interface of MnGa grown epitaxially on GaAs have been demonstrated.[20,21,22] Moreover, the MnGa alloy has large perpendicular magnetic anisotropy[23] and high Curie temperature (~ 600 K), which is necessary for thermal stability in nanoscale devices.

Recently, we prepared a lateral spin-valve device with MnGa electrodes and a 600 nm-long GaAs channel, and obtained a large MR ratio up to 12% and a spin-dependent output voltage of 33 mV, which are much higher than those in lateral FM/SC/FM spin-valve devices.[24] In this work, we investigated its spin-dependent transport properties in more details. We show that the observed large

MR originates from the intrinsic spin-valve effect rather than parasitic effects. Our results are an important step toward realization of practical spin transistors.

## 2. Epitaxial growth and device structure

The spin-valve device is composed of $Mn_{0.6}Ga_{0.4}$ (10 nm) / GaAs:Se (20 nm) with an electron density of $n \sim 10^{18}$ cm$^{-3}$, grown by molecular beam epitaxy (MBE) on a semi-insulating GaAs(001) substrate. The MBE growth process is as follows.[20] First, we removed the surface oxide layer of the GaAs substrate by thermal annealing at 580°C, and grew 20 nm thick Se-doped degenerated GaAs channel layer. The substrate was cooled down to room temperature for the deposition of 4 monolayers Mn–Ga–Mn–Ga template. Then, the substrate was heated to 250°C for the deposition of a MnGa thin film with the total thickness of 10 nm. During the MBE growth, we used reflection high-energy electron diffraction (RHEED) to monitor the crystallinity and the surface morphology of the samples. Figure 1(a) and (b) show the RHEED patterns observed along the GaAs [1̄10] direction after the growth of the GaAs:Se channel layer and MnGa layer, respectively. Observed streaky patterns indicate the high quality of the MnGa film. After epitaxial growth, a 600 nm-long GaAs channel was formed by using electron beam lithography (EBL) and Ar ion milling to separate the source and drain MnGa electrodes, as schematically shown in Fig. 1 (c). First, we used EBL, electron-beam evaporation and lift-off techniques to pattern a 20 nm-thick Au / 5 nm-thick Cr hard mask with a 600 nm-long and 150 nm-wide gap. Then, the exposed MnGa area was etched away by Ar ion milling to define a 600 nm GaAs channel. Figure 1(d) shows the top-view scanning electron microscopy (SEM) image of our lateral spin-valve device. A GaAs channel with a length of 600 nm was formed between the MnGa electrodes. Although the shape of the source and drain MnGa electrodes appear symmetric, their coercive forces near the GaAs channel are different due to the following reasons. The coercive force is determined by the domain wall pinning energy near the channel, which is affected by the damage caused by Ar ion milling. Since ion milling damage is not exactly the same but slightly different between the source and drain electrodes, it causes a difference in the domain wall pining energy and the coercive force. In fact, we found that the coercive force differs by 200 Oe, as shown later. Therefore, the spin-valve effect can be observed in this structure. In addition, large MR can be expected by spin injection using tunneling transport through the MnGa/GaAs Schottky barrier.[25]

## 3. Spin dependent transport properties

Conventionally, Hanle effect measurements and 4-terminal non-local measurements are effective methods to accurately evaluate the spin injection efficiency.[26,27,28] However, those methods require micron-length channel for spin precession, which is not the topic of this work. Here, we used the 2-terminal local spin-valve measurement, which is fundamentally important for the operation of spin transistors. A drawback of this method is that it suffers from parasitic local effects, such as the anisotropic magnetoresistance (AMR) effect of the ferromagnetic electrodes[29] and the tunneling anisotropic magnetoresistance (TAMR) effect at ferromagnet/semiconductor interface.[30,31] However, we emphasize here that, in principle, the AMR or TAMR effect does not occur in our case of MnGa electrodes with perpendicular magnetization due to the Onsager reciprocal principle. The AMR or TAMR effect requires the change of the relative angle between local magnetic moments and the current direction (for AMR) or a particular crystal axis (for TAMR). Such a situation is possible for the case of in-plane magnetization reversal when the local magnetic domains can rotate by 90 degree. Since our MnGa films have strong perpendicular magnetic anisotropy, the magnetization reversal of the MnGa electrodes occurs by domain wall movement between upward and downward magnetic domains. Thus, the local magnetic moment direction only changes from upward to downward or vice versa, and it is always perpendicular to the film plane during the magnetization reversal process. According to the Onsager reciprocal principle, the resistance should be the same for the upward and downward magnetic domains. Thus, no AMR or TAMR effect can be expected during the magnetization reversal of a perpendicular magnetic layer.

Figure 2 shows the drain-source current $I_{DS}$ – drain-source voltage $V_{DS}$ characteristics at various temperatures. The inset shows the drain-source resistance $R_{DS}$ as a function of temperature measured at $V_{DS}$ = 100 mV. The $I_{DS}$ – $V_{DS}$ characteristics shows Ohmic behavior above 100 K, and $R_{DS}$ is almost unchanged down to 100 K. Below 50 K, the $I_{DS}$ – $V_{DS}$ characteristics is non-linear, and $R_{DS}$ keeps increasing with decreasing temperature down to 4 K, indicating that thermionic emission of electrons over the MnGa/GaAs Schottky barrier is suppressed and tunnel current through the Schottky barrier becomes dominant. From these data, we can show that even if we assume the AMR effect existed in the MnGa electrodes, the contribution of this effect would be negligible. The contribution of the parasitic AMR to the total spin-valve effect can be estimated as follows. $R_{DS}$ can be decomposed into two components: $R_{DS} = R_{MnGa} + R_{SV}$, where $R_{MnGa}$ is the parasitic series resistance of the MnGa electrodes and $R_{SV}$ is the intrinsic resistance of the spin-valve structure. Since the parasitic $R_{MnGa}(T)$ decreases with lowering temperature $T$, $R_{MnGa}(T) < R_{MnGa}(300\ K) < R_{DS}(300\ K) = 144\ k\Omega$. Assuming that the AMR effect of $\Delta R_{MnGa}/R_{MnGa}$ is in the order of 1%, the parasitic resistance change of MnGa $\Delta R_{MnGa}$ is in order of 1 k$\Omega$. Since $R_{DS}\ (4\ K) = 1.1\ M\Omega$, the parasitic AMR effect of $\Delta R_{MnGa}/R_{DS}$ at 4 K should be in the order of 0.1%, which is too small to account for the observed MR ratio of 12%, as shown later in Fig. 3.

Figure 3 shows the representative MR curves obtained at 4 K with various $V_{DS}$ = 150, 200, 300, and 400 mV, while applying a magnetic field $H$ perpendicular to the film plane. In Fig. 3(a)–(d), the red and blue curves correspond to major loops obtained by sweeping the magnetic field $H$ from +1.5 kOe to -1.5 kOe and from -1.5 kOe to +1.5 kOe, respectively. Clear magnetoresistance in the major loops were observed. Importantly, a large MR ratio of 12% was obtained at $V_{DS}$ = 150 mV, which is much higher than those reported for lateral FM/SC/FM spin-valve devices. Clear minor loops (green curves) were also observed, indicating a difference of 200 Oe for the coercive force between the source and drain electrodes.

The large MR ratios obtained in our spin-valve device ensure that the parasitic AMR effect is not the origin. To further confirm that, we investigated the resistance change $\Delta R_{DS}$ between parallel and antiparallel magnetization at various $V_{DS}$, which are shown in Fig. 4(a). Since the parasitic $R_{MnGa}$ of MnGa electrodes does not depend on the bias voltage, the resistance change $\Delta R_{MnGa}$ due to the AMR effect of MnGa should not depend on $V_{DS}$. By contrast, the resistance change $\Delta R_{SV}$ due to the spin-valve effect depends on $V_{DS}$ because the spin injection efficiency depends on the bias voltage. In Fig. 4(a), $\Delta R_{DS}$ decreases with increasing $V_{DS}$, indicating that the observed MR signals do not originate from the AMR effect but the spin-valve effect.

Figure 4(b) shows the $V_{DS}$ dependence of the MR ratio (= $\Delta R_{DS}/R_{DS}$). The spin-valve effect was observed up to at least 1 V. The voltage $V_{half}$ at which the spin-valve ratio is reduced by half is about 300 mV, which is comparable to that of magnetic tunnel junctions and much larger than that of GaMnAs-based spin-valve devices. This demonstrates the advantage of our spin-valve structure. Finally, we report the spin-dependent performance of our device. For practical applications, large spin-dependent output voltage $\Delta V = \Delta R_{DS}/R_{DS}$ of the order of 100 mV is required for correct read-out of the embedded magnetic memory data. Figure 4(c) shows the $V_{DS}$ dependence of the spin-dependent output voltage $\Delta V$. In our device, $\Delta V$ = 33 mV was obtained at $V_{DS}$ = 1 V, which is the highest value reported in lateral FM/SC/FM spin-valve devices. However, thermal emission of electrons over the Schottky barrier prevents observation of the spin-valve effect at higher temperatures, due to the low Schottky barrier height (which is estimated to be several tens of meV). This problem can be solved by inserting an AlAs tunnel barrier between the MnGa and the GaAs:Se channel, or by using the ballistic transport when the channel length is sufficiently shorter than the electron mean free path and the electron scattering probability in the channel is negligibly small. In our previous studies of nano-scale Si-based spin-valve devices, the spin-valve effect was obtained by using (quasi) ballistic transport even at room temperature.[15,19] By downsizing the GaAs channel to sub-100 nm, it may be possible to obtain spin-valve signals at room temperature.

## 4. Conclusions

We have prepared a lateral spin-valve device with perpendicularly magnetized MnGa electrodes and a 600 nm-long GaAs channel and investigated its spin-dependent transport properties. We obtained large MR ratios up to 12% and large spin-dependent output voltages up to 33 mV. These values are the highest in lateral FM/SC/FM spin-valve devices. The $V_{DS}$ dependence of the MR indicates that observed MR signal originates from the spin-valve effect. Our results provide an important step towards realization of practical spin transistors.

**Acknowledgements**

This work was partly supported by Grants-in-Aid for Scientific Research (No. 16H02095, No. 18H03860), the CREST program of the Japan Science and Technology Agency (JPMJCR1777), the Nanotechnology platform 12025014, and the Spintronics Research Network of Japan (Spin-RNJ).

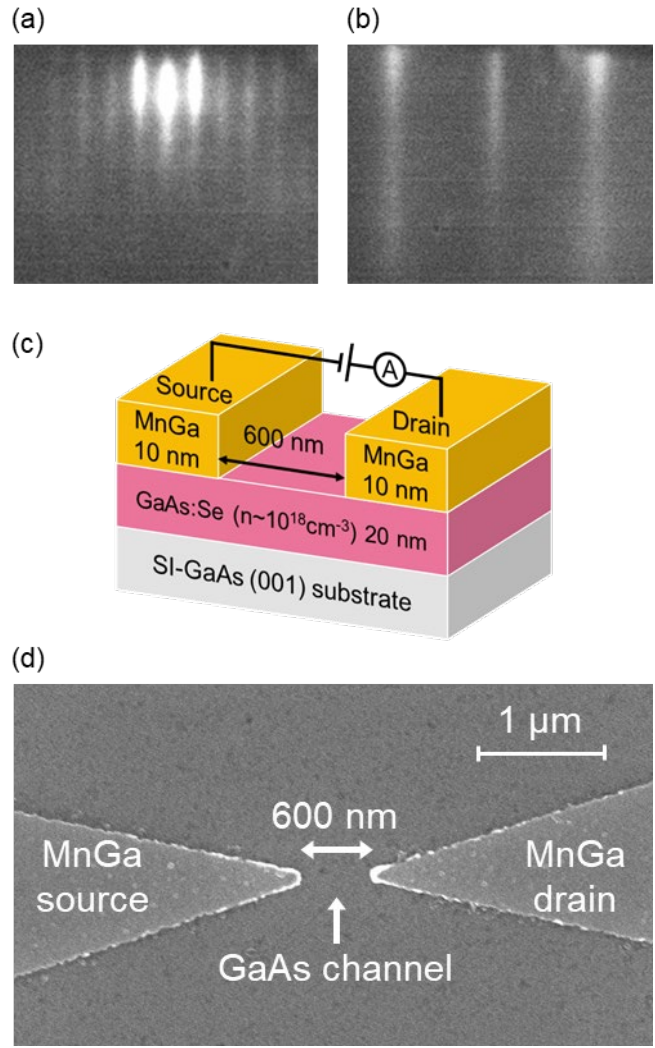

**Figure 1 (a), (b)** RHEED patterns taken along the [110] direction after growth of the GaAs:Se channel layer and MnGa layer, respectively. **(c)** Schematic structure of the lateral spin-valve device with MnGa electrodes and a 600 nm-long GaAs channel. **(d)** Scanning electron microscopy image (top view) of a device. A GaAs channel with a length of 600 nm was formed between the MnGa electrodes.

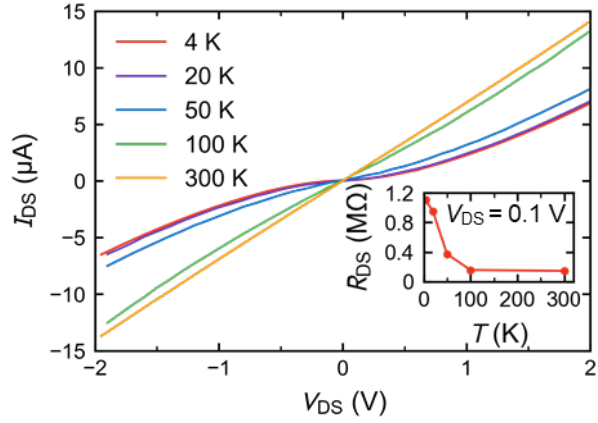

**Figure 2.** Drain-source current $I_{DS}$ – drain-source voltage $V_{DS}$ characteristics of a MnGa/GaAs/MnGa lateral spin-valve device at various temperatures ($T$ = 4 – 300 K). The $I_{DS}$ – $V_{DS}$ curve shows nonlinear behavior below 50 K. The inset shows the drain-source resistance $R_{DS}$ as a function of temperature measured at $V_{DS}$ = 100 mV.

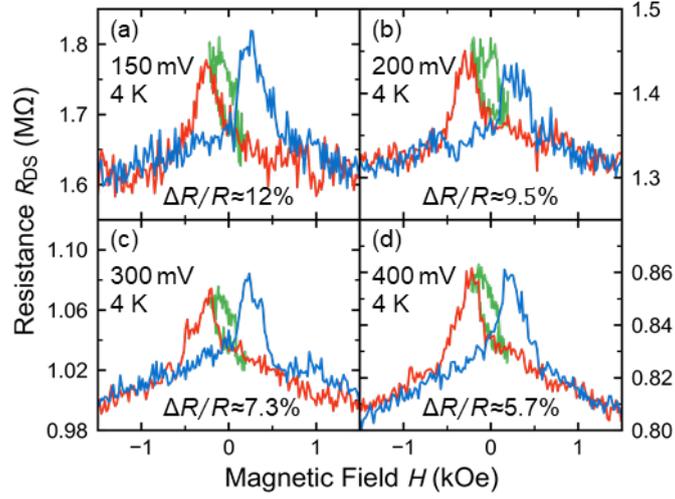

**Figures 3 (a)–(d)** Magnetoresistance (MR) characteristics obtained at 4 K with $V_{DS}$ = 150, 200, 300, and 400 mV, respectively, while applying a magnetic field $H$ perpendicular to the film plane. The red and blue curves correspond to major loops obtained by sweeping $H$ from +1.5 kOe to -1.5 kOe and from -1.5 kOe to +1.5 kOe, respectively. The green curves are minor loops. The MR ratio reaches 12% at $V_{DS}$ = 150 mV.

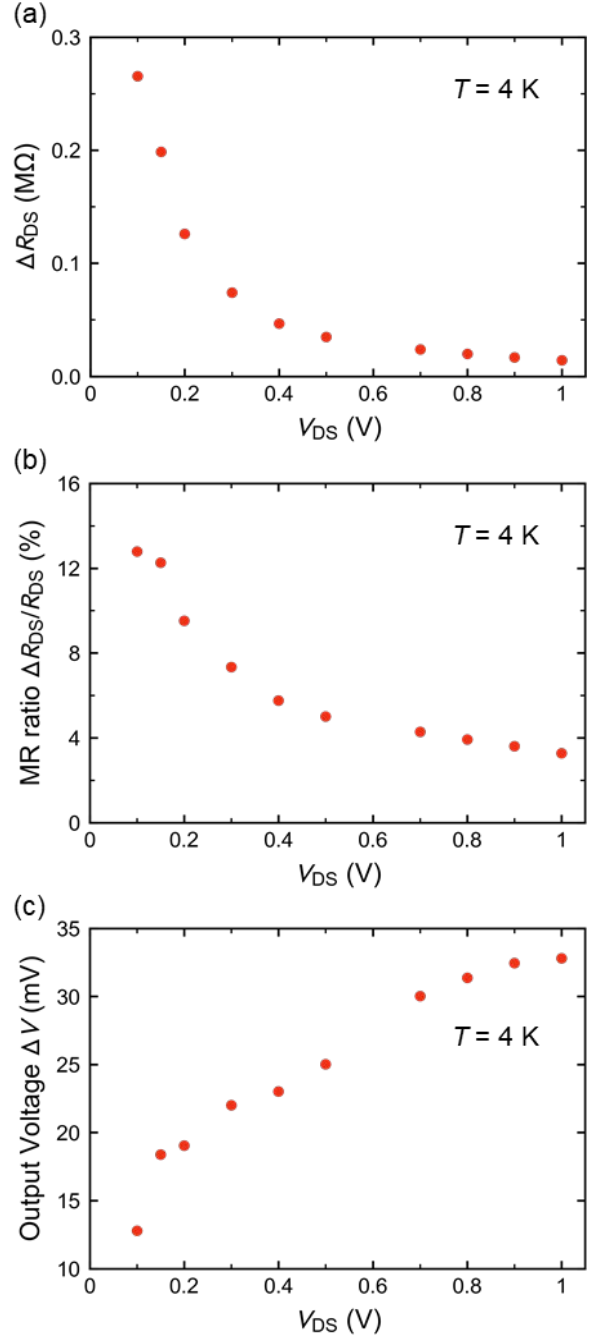

**Figure 4 (a)** Resistance change $\Delta R_{DS}$ as a function of $V_{DS}$. **(b)** $V_{DS}$ dependence of the MR ratio = $\Delta R_{DS}/R_{DS}$. **(c)** $V_{DS}$ dependence of the spin-dependent output voltage $\Delta V = (\Delta R_{DS}/R_{DS})V_{DS}$. The highest spin-dependent output voltage of 33 mV was obtained at $V_{DS}$ = 1 V.

**References**

[1] S. Sugahara and M. Tanaka, Appl. Phys. Lett. **84**, 2307 (2004).

[2] M. Tanaka and S. Sugahara, IEEE Trans. Electron Devices **54**, 961 (2007).

[3] S. Sugahara and M. Tanaka, J. Appl. Phys. **97**, 10D503 (2005).

[4] T. Matsuno, S. Sugahara, and M. Tanaka, Jpn. J. Appl. Phys. **43**, 6032 (2004).

[5] R. Nakane, T. Harada, K. Sugiura, and M. Tanaka, Jpn. J. Appl. Phys. **49**, 113001 (2010).

[6] T. Sasaki, Y. Ando, M. Kameno, T. Tahara, H. Koike, T. Okawa, T.Suzuki, and M. Shiraishi, Phys. Rev. Appl. **2**, 034005 (2014).

[7] T. Sasaki, T. Suzuki, Y. Ando, H. Koike, T. Oikawa, Y. Suzuki, and M. Shiraishi, Appl. Phys. Lett. **104**, 052404 (2014).

[8] T. Tahara, H. Koike, M. Kameno, T. Sasaki, Y. Ando, K. Tanaka, S. Miwa, Y. Suzuki, and M. Shiraishi, Appl. Phys. Express **8**, 113004 (2015).

[9] S. Sato, R. Nakane, T. Hada, and M. Tanaka, Phys. Rev. B 96, 235204 (2017).

[10] M. Oltscher, F. Eberle, T. Kuczmik, A. Bayer, D. Schuh, D. Bougeard, M. Ciorga and D. Weiss, Nat. Commun. **8**, 1807 (2017).

[11] H. Asahara, T. Kanaki, S. Ohya, and M. Tanaka, Appl. Phys. Express **11**, 033003 (2018).

[12] L. Chen, X. Yang, F. Yang, J. Zhao, J. Misuraca, P. Xiong, and S. von Molnar, Nano Lett. **11**, 2584 (2011).

[13] G. Schmidt, D. Ferrand, L. W. Molenkamp, A. T. Filip, and B. J. van Wees, Phys. Rev. B **62**, R4790 (2000).

[14] K. Natori, Appl. Surf. Sci. **254**, 6194–6198 (2008).

[15] D. D. Hiep, M. Tanaka, and P. N. Hai, Appl. Phys. Lett. **109**, 232402 (2016).

[16] D. D. Hiep, M. Tanaka, and P. N. Hai, J. Appl. Phys. **122**, 223904 (2017).

[17] D. D. Hiep, M. Tanaka, and P. N. Hai, Adv. Nat. Sci: Nanosci. Nanotechnol. **10**, 025001 (2019).

[18] D. Saha, M. Holub, P. Bhattacharya, and Y. C. Liao, Appl. Phys. Lett. **89**, 142504 (2006).

[19] P. N. Hai, Y. Sakata, M. Yokoyama, S. Ohya, and M. Tanaka, Phys. Rev. B **77**, 214435 (2008).

[20] M. Tanaka, J. P. Harbison, J. DeBoeck, T. Sands, B. Philips, T. L. Cheeks, and V. G. Keramidas, Appl. Phys. Lett. **62**, 1565 (1993).

[21] L.J. Zhu, S.H. Nie, K.K. Meng, D. Pan, J.H. Zhao, H.Z. Zheng, Adv. Mater. **24**, 4547 (2012).

[22] L. J. Zhu, D. Pan, S. H. Nie, J. Lu, and J. H. Zhao, Appl. Phys. Lett. **102**, 132403 (2013).

[23] A. Sakuma, J. Magn. Magn. Mater. **187**, 105 (1998).

[24] K. Chonan, N. H. D. Khang, M. Tanaka, and P. N. Hai, SSDM 2019 Extended Abstract, E-5-04 (2019).

[25] G Schmidt and L W Molenkamp, Semicond. Sci. Technol. **17**, 310 (2002).

[26] X. Lou, C. Adelmann, S. A. Crooker, E. S. Garlid, J. Zhang, S. M. Reddy, S. D. Flexner, C. J. Palmstrøm, and P. A. Crowell, Nat. Phys. **3**, 197 (2007).

[27] B. Huang, D. J. Monsma, and I. Appelbaum, Phys. Rev. Lett. **99**, 177209 (2007).

[28] R. Jansen, S. P. Dash, S. Sharma, and B. C. Min, Semicond. Sci. Technol. **27**, 083001 (2012).

[29] T. McGuire and R. Potter, IEEE Trans. Magn. **11**, 1018 (1975).

[30] C. Gould, C. Rüster, T. Jungwirth, E. Girgis, G. M. Schott, R. Giraud, K. Brunner, G. Schmidt, and L. W. Molenkamp, Phys. Rev. Lett. **93**, 117203 (2004).

[31] C. Rüster, C. Gould, T. Jungwirth, J. Sinova, G. M. Schott, R. Giraud, K. Brunner, G. Schmidt, and L. W. Molenkamp, Phys. Rev. Lett. **94**, 027203 (2005).